\RequirePackage{fix-cm}
\documentclass[aps,prd,showpacs,showkeys,superscriptaddress,twocolumn]{revtex4-1}
\usepackage[colorlinks,linkcolor=blue,anchorcolor=blue,citecolor=blue,urlcolor=blue,breaklinks=true]{hyperref}
\usepackage[T1]{fontenc}
\usepackage{graphicx}
\usepackage{amsmath}
\usepackage{amssymb}
\usepackage{slashed}
\usepackage{latexsym}
\usepackage{epsfig}
\usepackage{amsbsy}
\usepackage{array}
\usepackage{setspace}
\usepackage{bm}
\usepackage{lipsum}
\usepackage{mathrsfs}
\usepackage{float}
\usepackage{changes}
\usepackage{color}
\usepackage{booktabs}
\usepackage{mathptmx}
\usepackage{graphicx}
\usepackage{epstopdf}
\usepackage{ulem}
\usepackage{mathtools}
\usepackage{extarrows}
\usepackage{subcaption}
\newcommand{\Lagr}{\mathcal{L}}

\begin{document}
	\author{Bing-Jun Zuo}
	\email{dg20220091@smail.nju.edu.cn}
	\affiliation{Department of Physics, Nanjing University, Nanjing 210093, China}
	\author{Zheng Zhang}
	\affiliation{Department of Physics, Nanjing University, Nanjing 210093, China}
	\author{Chao Shi}
	\email{cshi@nuaa.edu.cn}
	\affiliation{Department of Nuclear Science and Technology, Nanjing University of Aeronautics and Astronautics, Nanjing 210016, China}
	\author{Yong-Feng Huang}
	\email{hyf@nju.edu.cn}
	\affiliation{School of Astronomy and Space Science, Nanjing University, Nanjing 210023, China}
	\affiliation{Key Laboratory of Modern Astronomy and Astrophysics (Nanjing University), Ministry of Education, Nanjing 210023, China}
	\affiliation{Xinjiang Astronomical Observatory, Chinese Academy of Sciences, Urumqi 830011, China}

		\title{Strongly interacting matter in a sphere at nonzero magnetic field}
	
	\begin{abstract}
		We investigate the chiral phase transition within a sphere under a uniform background magnetic field.
		The Nambu--Jona-Lasinio (NJL) model is employed and the MIT boundary condition is imposed
        for the spherical confinement. Using the wave
		expansion method, the diagonalizable Hamiltonian and energy spectrum are derived for
		the system. By solving the gap equation in the NJL model, the influence of magnetic field on quark matter in a sphere is studied. It is found that inverse magnetic catalysis occurs at small radii,
                 while magnetic catalysis occurs at large radii. Additionally, both magnetic catalysis and inverse
                 magnetic catalysis are observed at the intermediate radii ($R\approx4$ fm). 
	\end{abstract}
	
	\flushbottom
	\maketitle
	\thispagestyle{empty}
	
\section{Introduction}

It is found that finite size effects can have significant impacts on the state of dense matter
and phase transitions for systems whose scale ranges between 2 fm and 10
fm \cite{Chernodub:2017mvp,Chen:2017xrj,sadooghi2021inverse,Palhares_2011,klein2017modeling}.
This help us better understand the properties of Quantum Chromodynamics (QCD) fireball produced in high energy heavy-ion collision
(HIC) experiments. Furthermore, the magnetic field strength ($B$) can
reach up to $10^{15}$ ---
$10^{18}G$ in the HIC  \cite{PhysRevC.83.054911,PhysRevC.85.044907}, which can influence the energy of the system and
even lead to the breaking or restoration of chiral symmetry \cite{miransky2015quantum}.

In most conditions, the presence of an external magnetic field can lead to the breakdown of $U(1)_A$ symmetry and dimensional reduction, resulting in an increase in chiral condensate as the magnetic field increases, which is known as magnetic catalysis \cite{PhysRevLett.73.3499,Ebert:1999ht,RevModPhys.88.025001}.
However, as argued by the authors of \cite{Bali:2012zg, Bali:2011qj},
it is also possible that the increase of the magnetic field at the critical
temperature ($T_c$) may lead to a decrease in the chiral condensate,
resulting in a reversal of the magnetic catalysis effect, referred to as
the inverse magnetic catalysis. The underlying reason for inverse
magnetic catalysis is still highly
debated \cite{Bruckmann:2013oba,Fraga:2012ev,Ferrer:2014qka,Cao:2021rwx}.
In this study, we will further investigate the phenomena of (inverse) magnetic catalysis
in a finite size system.  In this aspect, some
authors have recently studied  systems confined in a cylindrical
geometry \cite{chen2016analogy,sadooghi2021inverse}. Meanwhile, quark matter confined within a sphere can sometimes be a realistic condition. Therefore, we will consider a finite size system confined
in a spherical geometry in this study.

To calculate the effects of magnetic field on quark matter, QCD-like effective models are commonly used. Here we will employ
the Nambu--Jona-Lasinio (NJL) model, which has been widely employed to study the chiral
phase transitions \cite{buballa2005njl}. It is also important to consider
appropriate boundary conditions for a finite sphere.
There are various boundary conditions, such as the MIT
boundary condition \cite{PhysRevD.9.3471}, the periodic boundary condition \cite{Rajaraman:1982xf}, and the chiral condition \cite{zhang2020rotating}.
The MIT boundary condition originates from the MIT bag model, which is widely used for describing the strong interaction of
quarks. Here we will enforce the MIT boundary
condition to describe an impenetrable spherical cavity. Meanwhile, as an initial attempt,
we will consider the one-flavor NJL model as in \cite{sadooghi2021inverse} for simplicity,
and set the temperature and baryon chemical potential to zero.
 Yet, extending the 
study to two flavor case is also straightforward.

The results of this study may contribute to a better understanding and modeling of the behavior of quark-gluon matter in heavy-ion collisions,  where strong magnetic fields and finite size effects could play a role.

This paper is organized as follows. In Section \ref{s2}, the mode solutions
of free fermions in a finite sphere with the MIT boundary condition are
introduced. The solutions are then extended to the case with a strong
magnetic field by using the mode expansion method. In Section \ref{s3}, by solving the gap equation in the NJL model with the obtained modes, the fermion condensate and the relationship between the magnetic field and the effective mass of quark matter is obtained.
The occurrence of magnetic catalysis and inverse magnetic
catalysis at different radii is explored. Possible explanations for the
occurrence of inverse magnetic catalysis are suggested. 
Finally, we present our conclusions  in Section \ref{s4}.

	\section{Effects of magnetic field on fermions in a sphere}
	\label{s2}
	
	\subsection{Mode solutions in a sphere}
	
	Assuming that the magnetic field is uniform and is along the z-axis,
    the Dirac equation of fermions is	
	\begin{eqnarray}\label{dir}
		[i\gamma^{0}\partial_{t}&+i\gamma^{1}\left(\partial_x+iqeBy/2\right)
		+i\gamma^{2}\left(\partial_y-iqeBx/2\right) \notag
		\\&+i\gamma^{3}\partial_{z}-M]\psi=0,
	\end{eqnarray}
 where $e$ is the charge of electrons (in natural units, $e = 1/\sqrt{137}$), $q$ is the charge
    fraction of the fermions considered, and $qe$ is the charge of the fermions.
	Here, we use the Pauli-Dirac representation of the gamma matrices, i.e.
	\begin{equation}\gamma^{0}=\left(\begin{array}{cc}
			1 & 0 \\
			0 & -1
		\end{array}\right), \quad \gamma^{i}=\left(\begin{array}{cc}
			0 & \sigma_{i} \\
			-\sigma_{i} & 0
		\end{array}\right),
    \end{equation}
	where $\sigma_i$ are Pauli matrices. When the magnetic field strength is
	zero (i.e. $B=0$), the above equation has a normal solution
	of $\psi(x)=X u(x)e^{-iEt}$, where the constant $X$ can be determined by
	using the normalization condition. The free part of the Hamiltonian can
	be expressed as
	\begin{eqnarray}
		H_0=-i\gamma^{0}\gamma^{i}\partial_i+\gamma^{0}M.
	\end{eqnarray}
In the presence of a non-zero magnetic field ($B\neq0$), the total
Hamiltonian is given by $H=H_0+H'$, where
\begin{eqnarray}
	H' = \frac{qeB}{2}(\gamma^{0}\gamma^1y-\gamma^{0}\gamma^2x).
\end{eqnarray}
In our study, $H'$ is expressed in spherical coordinates as
\begin{eqnarray}
	H' = \frac{iqeBr}{2}\left(\begin{array}{cc}
		0 & \bigtriangleup \\
		\bigtriangleup & 0
	\end{array}\right),
\end{eqnarray}
with
\begin{eqnarray}
	\bigtriangleup=\sin\theta\left(\begin{array}{cc}
		0 & e^{-i\varphi} \\
		-e^{i\varphi}\ & 0
	\end{array}\right).
\end{eqnarray}\\
	Following Ref. \cite{zhang2020rotating}, we first consider the mode solutions
	for $B=0$. In spherical coordinates, there are four commuting
	operators: $\{H_0,J^2,J_z,K\}$, where $J^2$ represents the total angular operator
	and $K$ is defined as
	\begin{equation}
		K=\gamma^0(L\cdot\Sigma+1).
	\end{equation}
	Here, $L$ represents the orbital angular momentum operator, and $\Sigma$ is
	defined as
	\begin{equation}
		\Sigma=\left(\begin{array}{cc}
			\sigma & 0 \\
			0 & \sigma
		\end{array}\right).
	\end{equation}

The eigenstate can hence be characterized by the eigenvalues of the four operators,
namely $\{E,j(j+1),m_j,\kappa\}$. For convenience, we will
use $k=(E,j,m_j,\kappa)$ to denote each eigenstate from now on.
 The solution of the Dirac equation in spherical
coordinates reads
	\begin{equation}\label{solution}
		\begin{aligned}
			u_k(r,\theta,\varphi) =\left\{\begin{array}{cl} &\left[\begin{array}{cc}{\sqrt{\frac{E+M}{2E}}j_{j-\frac{1}{2}}(pr)\chi^+_{jm_j}} \\ {i\frac{E}{|E|}}\sqrt{\frac{E-M}{2E}}{j_{j+\frac{1}{2}}(pr)\chi^-_{jm_j}}\end{array}\right], \ \ \ \kappa>0,\\
			&\left[\begin{array}{cc}{\sqrt{\frac{E+M}{2E}}j_{j+ \frac{1}{2}}(pr)\chi^-_{jm_j}}\\{-i\frac{E}{|E|}}\sqrt{\frac{E-M}{2E}}{j_{j-\frac{1}{2}}(pr)
					\chi^+_{jm_j}}\end{array}\right], \ \ \ \kappa<0,\end{array}\right.
		\end{aligned}
	\end{equation}
	with $j=\frac{1}{2} , \frac{3}{2},\cdots$, $m_j=-j,-j+1,\cdots,j$,
	and $\kappa=\pm(j+\frac{1}{2})$.

	The function $\chi^\pm_{jm_j}$ in Eq. (\ref{solution}) is the spherical
	harmonic function, which is expressed as
	\begin{equation}\label{chi}
		\begin{aligned}
			&\chi_{j m_j}^{+}=\left(\begin{array}{l}
				\sqrt{\frac{j+m_j}{2 j}} Y_{j-\frac{1}{2}}^{m_j-\frac{1}{2}} \\
				\sqrt{\frac{j-m_j}{2 j}} Y_{j-\frac{1}{2}}^{m_j+\frac{1}{2}}
			\end{array}\right),\\
			&\chi_{j m_j}^{-}=\left(\begin{array}{c}
				\sqrt{\frac{j-m_j+1}{2(j+1)}} Y_{j+\frac{1}{2}}^{m_j-\frac{1}{2}} \\
				-\sqrt{\frac{j+m_j+1}{2(j+1)}} Y_{j+\frac{1}{2}}^{m_j+\frac{1}{2}}
			\end{array}\right).
		\end{aligned}
	\end{equation}

In this work, we adopt the MIT boundary condition, which requires that the
quark current vanishes on the boundary surface. Such a boundary condition
can be equivalently written as \cite{greiner2007quantum}
	\begin{equation}\label{cg}
		i\slashed{n}\psi(R)=\psi(R),
	\end{equation}
where $R$ represents the radius of the boundary surface
and $\slashed{n}=\gamma^\mu n_\mu$, with $n_\mu$ being the normal to
the boundary. Using the boundary condition of Eq.~(\ref{cg}), the allowed modes should satisfy
	\begin{equation}\label{eigenp}
		j_{l_\kappa}(pR)=\operatorname{sgn}(\kappa)\frac{ p}{E+M}j_{\overline{l}_\kappa}(pR),
	\end{equation}
	with
	\begin{equation}
		\begin{aligned}
			&l_{\kappa}=\left\{\begin{array}{cl}{\kappa-1}& {\text { for } \kappa>0} \\
				{-\kappa} & {\text { for } \kappa<0}\end{array}\right.,\\
			& \overline{l}_{\kappa}=\left\{\begin{array}{cc}{\kappa} & {\text { for } \kappa>0} \\
				{-\kappa-1} & {\text { for } \kappa<0}\end{array}\right..
		\end{aligned}
	\end{equation}
	Here $j_n$ is the $n$th spherical Bessel function.
According to  Eq.~(\ref{eigenp}), the momentum $p$ gets discretized, and we label the
i-th solution of Eq.~(\ref{eigenp}) as $p_{j\kappa,i}$. Using the on-shell condition,
the energy $E$ can be calculated from the momentum $p_{j\kappa,i}$, thus the label $k$ can
be equivalently written as
\begin{equation}\label{relab}
	k=(i,j,m_{j},\kappa).
\end{equation}

Imposing the normalization condition, the normalization constant of the Dirac wave function in Eq.~(\ref{solution})  is
\begin{equation}\label{nomc}
	\begin{aligned}	
		X_k=\left\{\begin{array}{cl} \frac{\sqrt{2}}{R|j_{j+\frac{1}{2}}(p_{j\kappa i}R)|}\sqrt{\frac{E+M}{2ER-(2j+1)+ \frac{M}{E}}},\ \ \kappa>0,\\
		\frac{\sqrt{2}}{R|j_{j-\frac{1}{2}}(p_{j\kappa i}R)|}\sqrt{\frac{E+M}{2ER+(2j+1)+ \frac{M}{E}}},\ \ \kappa<0.\end{array}\right.
	\end{aligned}
\end{equation}
	The mode solutions for a fermion under the MIT boundary condition can be summarized as
\begin{equation}
	U_k(t,r,\theta,\phi)=X_ku_k(r,\theta,\phi)e^{-iEt}.
\end{equation}
Consequently, the solutions for the anti-fermion can be obtained via charge conjugation
\begin{equation}
	V_k(t,r,\theta,\phi)=i\gamma^2U^*_k(t,r,\theta,\phi).
\end{equation}

\subsection{Mode solutions at nonzero magnetic field}
	
	We now extend our analysis to take the  magnetic field into account.
Analytical solutions of the Dirac equation in the presence of a magnetic field
are not directly available. Although perturbation methods can be used in certain
cases, they are less suitable for strong magnetic fields.
Here we go beyond the perturbative method and directly diagonalize the Hamiltonian. It can
potentially yield a more accurate result for the system
with a strong magnetic field. Such a method has been adopted in other studies, such as the electronic structures in a magnetic field of a spherical quantum dot \cite{wu2012electronic}.
	
   In the presence of a magnetic field, only $m_j$ and the 
        energy $E$ remain to be good quantum numbers.  
    We denote the particle eigenstate as
	$S_{m_j,n}$, which is now characterized by $m_j$ and a new label $n$
	for the energy eigenvalue. It can be expressed using a complete set of
	orthonormalized particle basis functions \{$U_k$\} as
	\begin{equation}\label{exp1}
	S_{m_j,n}=\sum_{ij\kappa} c_{ij\kappa,n}U_k.
\end{equation}
Note $k=(i,j,m_{j},\kappa)$ as  defined in Eq.~(\ref{relab}).
Correspondingly, the anti-particle eigenstate ($T_{m_j\kappa,n}$) is
\begin{equation}\label{exp2}
	T_{m_j,n}=\sum_{ij\kappa} c^*_{ij\kappa,n}V_k.
\end{equation}
Replace $\psi$ in Eq.~(\ref{dir}) with $S_{m_j,n}$ and $T_{m_j,n}$,
we can get the secular equation for a certain $m_j$
\begin{equation}\label{sec}
	\mid H_{ij\kappa,i'j'\kappa'}^{m_j}-E_n^{m_j}\delta_{ii'}\delta_{jj'}\delta_{\kappa\kappa'}\mid=0.
\end{equation}

The matrix elements $H_{ij\kappa,i'j'\kappa'}^{m_j}$ can be calculated as
\begin{widetext}
\begin{equation}\label{hnj}
	\begin{aligned}
		&H_{ij\kappa,i'j'\kappa'}^{m_j} \equiv \langle j',m_j,\kappa',i'\mid H\mid j,m_j,\kappa,i\rangle\\
		&=E_k\delta_{ii'}\delta_{jj'}\delta_{kk'}+\frac{qeBr}{2}\left[  A_1\frac{4jm_j+2m_j}{2j(2j+2)}\delta_{jj'}\delta_{kk'}+A_2\left( \frac{\sqrt{(j+m_j)(j-m_j)}}{2j}\delta_{j'(j-1)}+\frac{\sqrt{(j+m_j+1)(j-m_j+1)}}{2j+2}\delta_{j'(j+1)}\right) \right] ,
	\end{aligned}
\end{equation}
\end{widetext}
where $E_k=\sqrt{p_{j\kappa,i}^2+M^2}$.   $A_1$ and $A_2$ are given by
\begin{widetext}
\begin{equation}
	\begin{aligned}
		 A_1&=\left\{\begin{array}{cl}-\frac{(E_k'+m)p\displaystyle\int_0^Rr^2j_{j'-\frac{1}{2}}(p'r)j_{j+\frac{1}{2}}(pr)dr+(E_k+m)p'\int_0^Rr^2j_{j'+\frac{1}{2}}(p'r)j_{j-\frac{1}{2}}(pr)dr}{R^2|j_{j'+\frac{1}{2}}(p'R)j_{j+\frac{1}{2}}(pR)|\sqrt{2E_{k'}^2R-E_{k'}+M}\sqrt{2E_k^2R-E_k+M}}  & {\text { for } \kappa,\kappa'>0}\\
				 {-\frac{(E_k'+m)p\displaystyle\int_0^Rr^2j_{j'+\frac{1}{2}}(p'r)j_{j-\frac{1}{2}}(pr)dr+(E_k+m)p'\int_0^Rr^2j_{j'-\frac{1}{2}}(p'r)j_{j+\frac{1}{2}}(pr)dr}{R^2|j_{j'-\frac{1}{2}}(p'R)j_{j-\frac{1}{2}}(pR)|\sqrt{2E_{k'}^2R+E_{k'}+M}\sqrt{2E_k^2R+E_k+M}}} & {\text { for  } \kappa,\kappa'<0}\end{array}\right. , \\ \\
		 A_2&=\left\{\begin{array}{cl}-\frac{(E_k'+m)(E_k+m)\displaystyle\int_0^Rr^2j_{j'-\frac{1}{2}}(p'r)j_{j+\frac{1}{2}}(pr)dr+pp'\int_0^Rr^2j_{j'+\frac{1}{2}}(p'r)j_{j-\frac{1}{2}}(pr)dr}{R^2|j_{j'+\frac{1}{2}}(p'R)j_{j-\frac{1}{2}}(pR)|\sqrt{2E_{k'}^2R-E_{k'}+M}\sqrt{2E_k^2R+E_k+M}}  & {\text { for } \kappa<0,\kappa'>0}\\
			 {-\frac{(E_k'+m)(E_k+m)\displaystyle\int_0^Rr^2j_{j'+\frac{1}{2}}(p'r)j_{j-\frac{1}{2}}(pr)dr+p'p\int_0^Rr^2j_{j'-\frac{1}{2}}(p'r)j_{j+\frac{1}{2}}(pr)dr}{R^2|j_{j'-\frac{1}{2}}(p'R)j_{j+\frac{1}{2}}(pR)|\sqrt{2E_{k'}^2R+E_{k'}+M}\sqrt{2E_k^2R-E_k+M}}} & {\text { for  } \kappa>0,\kappa'<0}\end{array}\right. .
	\end{aligned}
\end{equation}
\end{widetext}
Note we utilize the superscript symbol ($'$) to indicate the index of the bra vector.

In deriving Eq. (\ref{hnj}), we have used the  following formulas
\begin{equation}
	\begin{aligned}\label{y1}
		e^{-i\varphi}\sin\theta Y_{l,m}=&\sqrt{\frac{(l+1-m)(l+2-m)}{(2l+1)(2l+3)}}Y_{l+1}^{m-1} \\
		&-\sqrt{\frac{(l+m)(l+m-1)}{(2l+1)(2l-1)}}Y_{l-1}^{m-1},
	\end{aligned}
\end{equation}
\begin{equation}
	\begin{aligned}\label{y2}
		e^{i\varphi}\sin\theta Y_{l,m}=&-\sqrt{\frac{(l+m+1)(l+m+2)}{(2l+1)(2l+3)}}Y_{l+1}^{m+1}\\&+\sqrt{\frac{(l-m)(l-m-1)}{(2l+1)(2l-1)}}Y_{l-1}^{m+1}.
	\end{aligned}
\end{equation}
Eq. (\ref{y1}) can be found in Ref. \cite{cohen1986quantum}, 
   and Eq. (\ref{y2}) can be derived by applying complex conjugation to Eq. (\ref{y1}).

Finally, we can obtain the numerical solutions of the
energy eigenvalues and eigenstates by diagonalizing the Hamiltonian matrix in Eq.~(\ref{hnj}).
	
	\section{FERMION CONDENSATE AND CHIRAL PHASE TRANSITION UNDER MAGNETIC FILED}
	\label{s3}
	
	We use the NJL model to describe the interaction between quarks.
	The Lagrangian density representing  the NJL model is
	\begin{equation}\label{lar}
		\Lagr = \bar{\psi}(i\gamma^\mu\partial_\mu -\gamma^\mu qe A_\mu-m_0)\psi
		+ \frac{G}{2}[(\bar{\psi}\psi)^2+(\bar{\psi}i\gamma^5\psi)^2],
	\end{equation}
	where $G$ is the coupling constant, $m_0$ is the current quark mass, 
    and $qe$ is still the charge of the fermions.
	Here we take the chiral limit, i.e. $m_0=0$. Although the term of $F_{\mu \nu}F^{\mu \nu}$ 
    in Eq. (\ref{lar}) affects the pressure, it is not relevant to the chiral phase 
    transition. As a result, it is neglected in our calculations.
	
	According to the mean-field approximation, the gap equation can be derived
by considering the quark condensate as
	\begin{equation}\label{gap}
		M=-G\langle\bar{\psi}\psi\rangle.
	\end{equation}
	In order to calculate the quark condensate, we need to perform a second quantization.
    The fermion field operator can be expressed as
	\begin{equation}
		\psi=\sum_{\lambda}\left[S_{\lambda} {\bf b_{\lambda}}+T_{\lambda} {\bf d_{\lambda}^\dagger}\right]. 
	\end{equation}
	Here we use $\lambda=(m_j,n)$ to identify different states. $S_{\lambda}$ and $T_{\lambda}$ are wave functions  given by Eq. (\ref{exp1}) and Eq. (\ref{exp2}).	The operators $ {\bf b_{\lambda}}$ and ${\bf d_{\lambda}^{\dagger}}$ are annihilation and  creation operators, satisfying the
	canonical anti-commutation relations, 
	\begin{equation}
		\{{\bf b_{\lambda}}, {\bf b_{\lambda'}^{\dagger}}\}=\delta(\lambda,\lambda^{\prime}),~~	\{{\bf d_{\lambda}}, {\bf d_{\lambda'}^{\dagger}}\}=\delta(\lambda,\lambda^{\prime}).
	\end{equation}
	All other anti-commutation relations are zero. The vacuum
	state $\mid0\rangle$ is defined by
	\begin{equation}
		{\bf b_{\lambda}}|0\rangle= {\bf d_{\lambda}}|0\rangle=0.
	\end{equation}

	According to Ref.~\cite{1999Statistical,zhang2020rotating}, we have the following relations of 
	\begin{equation}
		\begin{aligned}
			\langle {\bf b_{\lambda}^{\dagger}} {\bf b_{\lambda^{\prime}}}\rangle&=\frac{1}{e^{\beta(E_{\lambda}-\mu)}+1}\delta(\lambda,\lambda^{\prime}),\\
			\langle {\bf d_{\lambda}} {\bf d_{\lambda^{\prime}}^{\dagger}}\rangle&=1-\langle {\bf d_{\lambda^{\prime}}^{\dagger}}{\bf d_{\lambda}}\rangle=\left( 1-\frac{1}{e^{\beta(E_{\lambda}+\mu)}+1}\right)  \delta(\lambda,\lambda^{\prime}),\\
			\langle {\bf b_{\lambda}^{\dagger}}{\bf d_{\lambda^{\prime}}^{\dagger}}\rangle&=\langle {\bf d_{\lambda}}{\bf b_{\lambda^{\prime}}}\rangle=0,\\
			\bar{V}_{\lambda}{V}_{\lambda}&=-\bar{U}_{\lambda}{U}_{\lambda}, 
		\end{aligned}
	\end{equation}
	where $E_{\lambda}$ represents the eigenenergy in the presence of a magnetic field. 
    Note that in the absence of a magnetic field, the eigenstate is denoted as $k=(i,j,m_{j},\kappa)$ and $k'=(i',j',m_{j},\kappa')$, whereas in the presence of a magnetic field, it is denoted as $\lambda=(m_j,n)$.
	
Substituting the above equations into $\langle\bar{\psi} \psi\rangle $, we have
	\begin{equation}\label{fpp}
		\langle\bar{\psi}\psi\rangle=-\sum_{\lambda}w(E_{\lambda})\sum_{k}\sum_{k'}|c_{ij\kappa,n}||c_{i'j'\kappa',n}|\bar{{{U}}}_{k}U_{k'},
	\end{equation}
	where
	\begin{equation}
		w(E_{\lambda})= 1-\frac{1}{e^{\beta\left({E_{\lambda}}+\mu\right)}+1}-\frac{1}{e^{\beta\left({E_{\lambda}}-\mu\right)}+1}.
	\end{equation}
	$\bar{{{U}}}_{k}U_{k'}$ can be derived from Eqs.~(\ref{solution}) and (\ref {nomc}), which depends on the coordinate $r$ inside the sphere.
	To consider the magnetic field's impact on the entire sphere, we can calculate
	the average value of $\bar{{{U}}}_{k}U_{k}$ as
	\begin{equation} \label{uu}
		\overline{	 \bar{{{U}}}_{k}U_{k'}}=\frac{1}{V}\int_V\bar{{{U}}}_{k}U_{k'}~ dV .
	\end{equation}

In our study, the quark condensate $\langle\bar{\psi}\psi\rangle_{\text{main}}$ is introduced, which is defined as
	\begin{equation}\label{bcon}
		\begin{aligned}
			\langle\bar{\psi}\psi\rangle_{\text{main}}=&-\sum_{\lambda}w(E_{\lambda})\sum_{ij}\frac{1}{V}|c_{ij\kappa,n}|^{2}\\
			&\frac{-\text{sgn}(\kappa)E_k+(2j+1)M+2E_kRM}{2E^2_kR-\text{sgn}(\kappa)(2j+1)E_k+M}.	
		\end{aligned}
	\end{equation}
	$\langle\bar{\psi}\psi\rangle_{\text{main}}$ can be calculated by taking terms in 
    index $k$ and $k'$ from $\langle\bar{\psi}\psi\rangle$ in Eq. (\ref{fpp}). 
    It is found that $\langle\bar{\psi}\psi\rangle$ is contributed mainly 
    by $\langle\bar{\psi}\psi\rangle_{\text{main}}$. With the help 
    of $\langle\bar{\psi}\psi\rangle_{\text{main}}$, we can analyze the 
    behavior of $\langle\bar{\psi}\psi\rangle$ in some simple cases. When $R$ is 
    large enough and $B$ is very small, $\langle\bar{\psi}\psi\rangle$ reduces 
    to the normal quark condensate (in an infinite space): the 
    coefficients $c_{ij\kappa,n}$  equal 1 
    and  $\langle\bar{\psi}\psi\rangle=\langle\bar{\psi}\psi\rangle_{\text{main}}$ (for $B=0$).  
    Meanwhile, when $R$ approaches infinity, the fraction term 
    in $\langle\bar{\psi}\psi\rangle_{\text{main}}$ approaches $\frac{M}{E_k}$.  
    Additionally, the summation in Eq.~(\ref{bcon}) will be replaced by an integral 
    over the momentum $p$. Thus quark condensate is reduced to
	\begin{equation}
		\langle\bar{\psi}\psi\rangle \xlongequal[R\rightarrow \infty]{B\rightarrow 0} 
        -\int\frac{d^3p}{(2\pi)^3}w(E_{k})\frac{M}{E_k}.
	\end{equation}

  The non-renormalizability of the NJL model
  requires that a regularization scheme should be applied.
  Here we use the three-momentum cutoff scheme. 
  Following \cite{chen2016analogy,Chen:2017xrj,sadooghi2021inverse}, 
  we take the cutoff momentum and the charge fraction as 
  \begin{equation}
  	\Lambda=1000\ \text{MeV}, \ \ \ \ \ \ q=1.
  \end{equation}
   In case of a zero temperature, the gap equation (\Ref{gap}) can be written as
	\begin{equation}\label{0gap}
		\begin{aligned}
	M = & \frac{1}{V}\sum_{\lambda}\varTheta(\Lambda-\bar{p})
      \sum_{k}\sum_{k'}|c_{ij\kappa,n}||c_{i'j'\kappa',n}|\overline{\bar{{{U}}}_{k}U_{k'}}, 
		\end{aligned}
	\end{equation}
	where $\varTheta$ is the Heaviside function and $\bar{p}$ is the expected 
    value of the momentum, $\langle \lambda |p|\lambda\rangle$. The 
    coefficients $c_{ij\kappa,n}$ can be determined by solving the secular 
    equation Eq.(\ref{sec}). Since we have 
	adopted a cutoff momentum of $\Lambda$, the dimension of the
	Hamiltonian matrix will be reduced from infinite to a finite number,
	thus making it numerically solvable.

\begin{figure}[h]
	\centering
	\begin{subfigure}{0.45\textwidth}
		\centering
		\includegraphics[width=\linewidth]{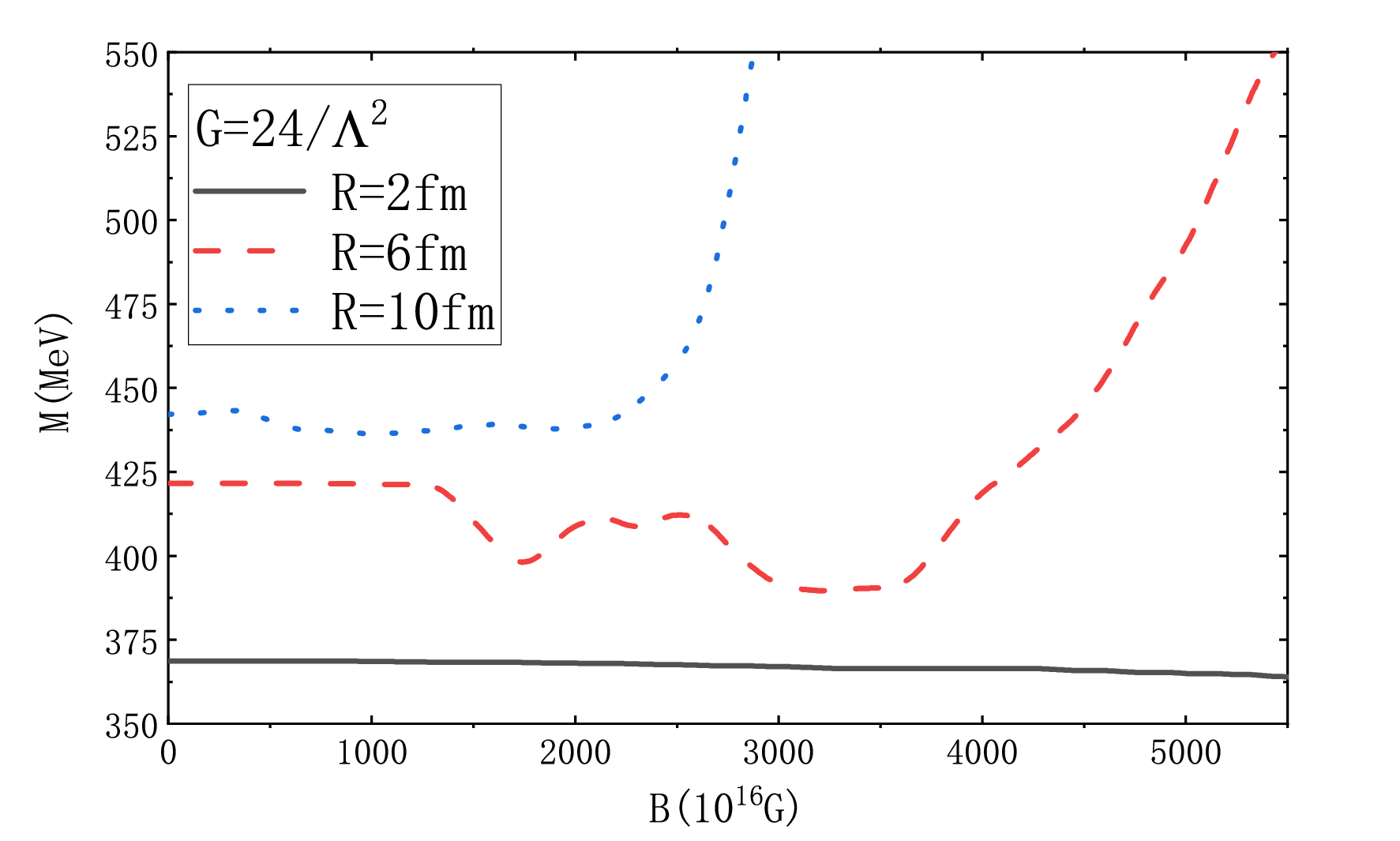}
		\caption{$G=24/\Lambda^2$}
		\label{fig:subfiga}
	\end{subfigure}
	\hfill
	\begin{subfigure}{0.45\textwidth}
		\centering
		\includegraphics[width=\linewidth]{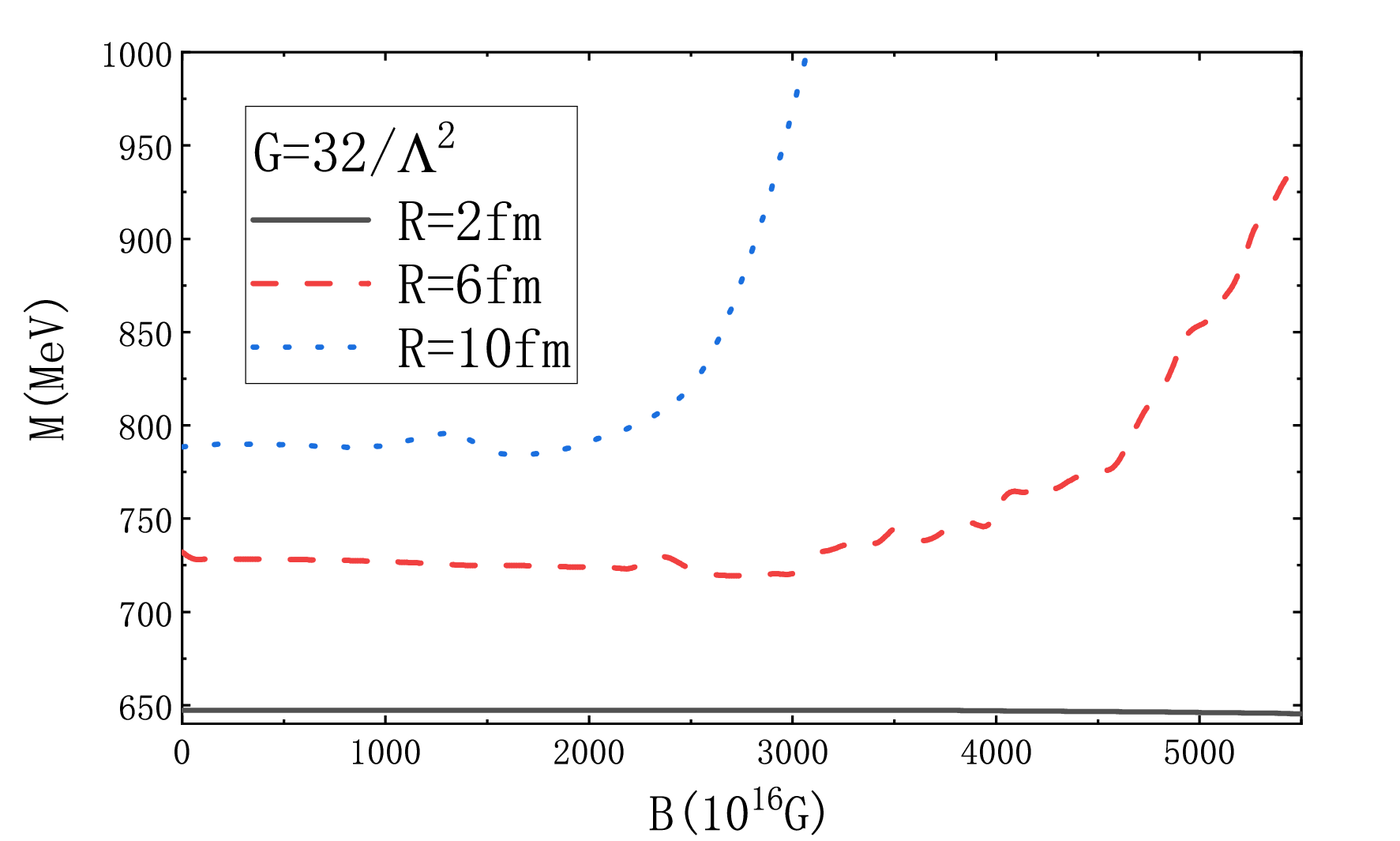}
		\caption{$G=32/\Lambda^2$}
		\label{fig:subfigb}
	\end{subfigure}
	\caption{Variation of the effective quark mass in a uniform magnetic field. The quarks
		are confined in a sphere with a radius of $R=2, 6, 10$ fm, respectively.
        Panels (a) and (b) correspond to $G=24/\Lambda^2$ and $G=32/\Lambda^2$, respectively. }
	\label{p1}
\end{figure}

	The effects of a strong magnetic field on the effective quark mass for quarks confined
	in a sphere are illustrated in Fig.~\ref{p1}.  We consider two different coupling 
    constants, $G=24/\Lambda^2$ and $G=32/\Lambda^2$ in Fig.~\ref{p1}. In both cases, 
    we see that the effective mass of quarks increases with the strength of the magnetic 
    field when $R$ is large. This phenomenon is exactly the so called chiral magnetic catalysis. 
    In general, such a chiral magnetic catalysis is also observed in the standard NJL model 
    in an infinite space \cite{GUSYNIN1996249,RevModPhys.88.025001,Menezes:2008qt}.   
    Moreover, when $R$ becomes small, the inverse magnetic catalysis is observed, i.e, 
    the effective mass of quarks decreases with the increase of the magnetic field.

\begin{figure}[ht]
	 	\centering\includegraphics[width=1\linewidth]{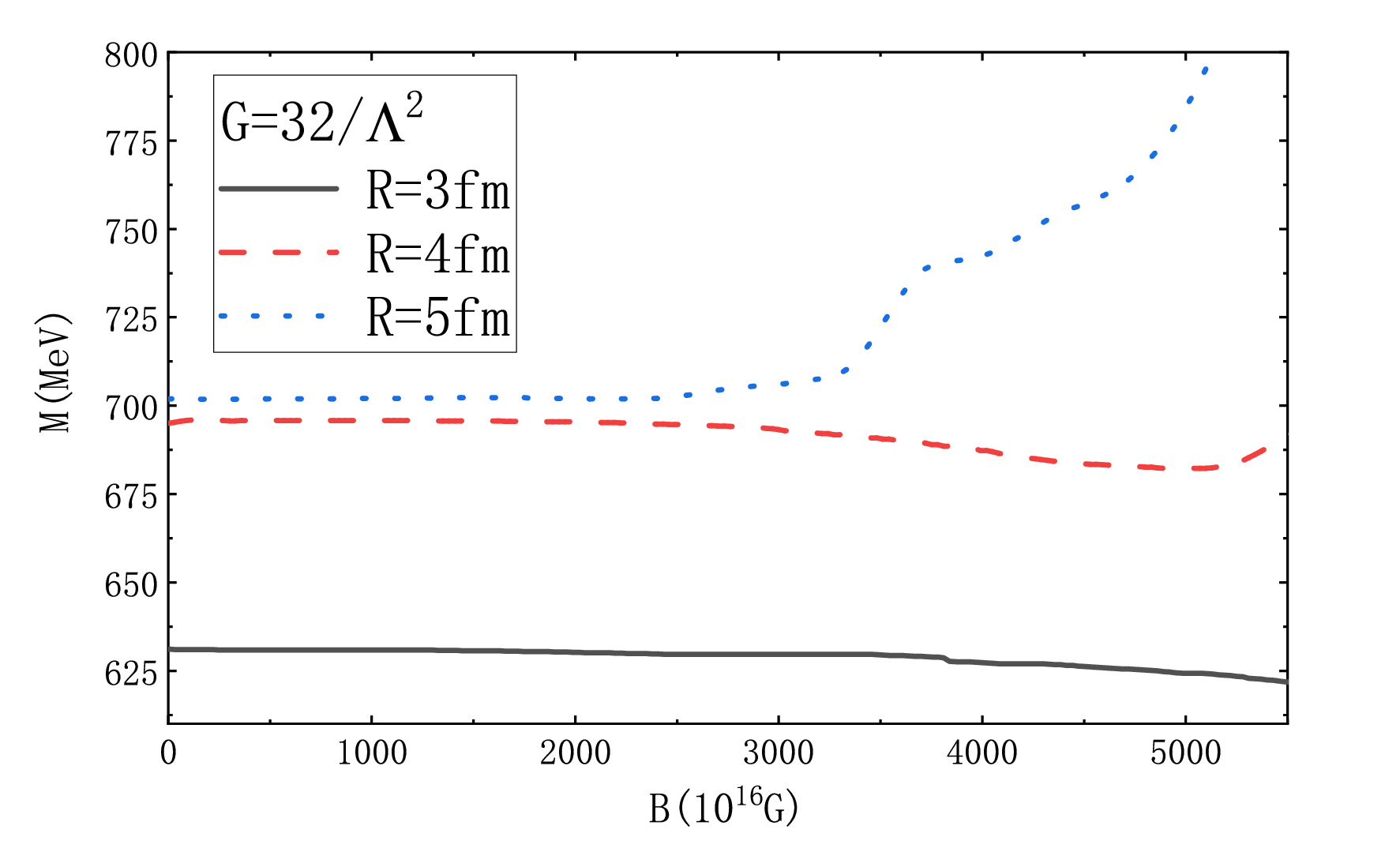}
	 	\caption{Variation of the effective quark mass in a uniform magnetic 
                  field near the critical radius for $G=32/\Lambda^2$.} 
	 	\label{p3}
	 \end{figure}

	The influence of the radius, $R$, on the occurrence of magnetic catalysis 
or inverse magnetic catalysis can be observed more clearly in Fig. \ref{p3}. 
Specifically, magnetic catalysis is observed at $R=5$ fm, inverse magnetic 
catalysis occurs at $R=3$ fm, and the case of $R=4$ fm falls between these 
two cases. It is well know that, at $T=0$, the standard NJL does
not exhibit inverse magnetic catalysis. To explore the potential causes of the inverse 
magnetic catalysis here, we should return to Eq. (\ref{bcon}) again. When $R$ is  
small, the contribution of the Lowest Orbital Level (LOL), i.e.,  $j=\frac{1}{2}$ and $\kappa>0$,  
reduces to
   	\begin{equation}\label{lol}
   	\langle\bar{\psi}\psi\rangle_{\text{main}} \xlongequal[\text{LOL}]{R\rightarrow 0}  
    -\sum_{\lambda}w(E_{\lambda})\sum_{i}\frac{1}{V}|c_{i,n}|^{2}
   	\frac{-E_k+2M}{-2E_k+M}.
   \end{equation}
Since $E_k=\sqrt{p_{j\kappa,i}^2+M^2}$, the contribution of these momentum modes 
to quark condensate will be positive for $p_{j\kappa,i}^2<\frac{M^2}{3}$, yielding 
an anomaly value. With the increase of the magnetic field, the energy gap between 
the orbital levels increases, causing the system to prefer the LOL and resulting 
in the inverse magnetic catalysis. Another factor is the intrinsic truncation caused 
by the small radius, which prevents the increase of Landau levels and the density 
of states. It further contributes to the inverse magnetic catalysis. 
	
Some oscillations could be seen in Fig. \ref{p1} when $R=6, 10$ fm. They are quite 
similar to the de Haas-van Alphen oscillations \cite{lifshitz2013statistical}. Such 
an oscillation behavior could be caused by the variation of the density of states 
due to the Landau quantization, which has also been observed in the standard NJL  
model \cite{PhysRevD.84.014016,Cao:2023bmk}. On the other hand, when the radius 
is small, the intrinsic truncation imposes a cutoff on the Landau levels, 
suppressing the variation of the density of states. Consequently, the de 
Haas-van Alphen oscillations cease at small radii.

	\section{ SUMMARY AND DISCUSSION }
	\label{s4}
	
In this study, quark matter confined in a sphere with a strong
uniform magnetic field is studied. The wave functions and energy
levels are solved by diagonalizing the Hamiltonian numerically, 
using eigen solutions of Hamiltonian with a zero magnetic field as 
basis. The NJL model is employed, and by solving its gap equation, 
the inverse magnetic catalysis effect and magnetic catalysis effect 
is studied  for a confined sphere  with various radii. It is found 
that when the radius of the sphere is large, magnetic catalysis 
occurs, whereas when the radius is small, inverse magnetic catalysis 
occurs. At the intermediate region ($R\approx4$ fm), both phenomena 
are present. It is argued that the inverse magnetic catalysis could be 
caused by the anomalous contribution from the LOL. Additionally, the 
intrinsic truncation due to small radius prevents the increase of 
Landau levels and density of states, which may also contribute to 
the inverse magnetic catalysis.

For simplicity, we mainly adopt the one-flavor NJL model with $q=1$ to investigate 
the chiral phase transition in our study. In fact, we have also performed 
calculations for the two-flavor NJL model, in which $q_u=\frac{2}{3}$ and 
$q_d=-\frac{1}{3}$. It is found that the results are generally similar. 
In the future, more realistic conditions should be considered. For example,
the study of finite temperature effects and the impact of nonzero current quark mass in a two-flavor model might be of special interest, which could 
provide additional insights into the behavior of QCD fireballs produced in 
heavy-ion collisions under realistic conditions.
	
\section*{Acknowledgements}
	
We thank Yong-Hui Xia for helpful discussions.
This study is supported by the National Natural Science Foundation of China
(Grant Nos. 12233002, 12041306), by the National Key R\&D Program
of China (2021YFA0718500), by National SKA Program of China No. 2020SKA0120300,
and by the Fundamental Research Funds for the Central Universities, NO. 1227050553.
YFH also acknowledges the support from the Xinjiang Tianchi Program.
	
	\bibliographystyle{apsrev4-1}
	\bibliography{MYREF}
	
\end{document}